# Plasmonic Induced Potential in Metal-Semiconductor Composits

*Shiva Shahin\*, Palash Gangopadhyay, and Robert A. Norwood*

*College of Optical Sciences, University of Arizona, Tucson, AZ, USA*

[*shshahin@ucsd.edu](mailto:shshahin@ucsd.edu)

**Abstract:** We utilize experimental tools such as conductive atomic force microscopy and electrostatic force microscopy on photo-excited arrays of gold nanoparticles on indium tin oxide substrate. The nanoparticles are partially covered by a thin semiconductor polymer. The change in the current and potential profiles of the composite metal-semiconductor sample after excitation at plasmonic resonance frequency of metallic nanoparticles is analyzed.

Plasmonic metallic nanostructures display remarkable optical properties that arise from their strong interaction with resonant photons through an excitation of surface plasmon resonance (SPR)[1–3]. When photo-excited, collective oscillations of valence electrons oscillate with the same frequency as the incident radiation and eventually enter resonance, giving rise to intense, highly localized electromagnetic fields[3,4]. After SPR excitations, plasmons can decay by radiating their energy resulting in light scattering[1,5] or decay nonradiatively by transferring the accumulated energy to electrons in the conduction band of the material producing hot electrons [6,7] or by converting absorbed light to heat[8,9]. Although this heating is generally regarded as a limitation for optical applications, recent work has demonstrated how a plasmonic resonance can act as a heat engine that uses thermal energy from the absorption of light to move electrons and produce static electric potentials[10].

Generally, in various applications, the optical properties of plasmonic nanostructures are exploited by putting them in contact with a semiconductor. However, the impact of plasmon-generated heat on the electrostatic potential of the composite metal-semiconductor structure has so far not been observed. In this work we utilize experimental tools such as conductive atomic force microscopy (C-AFM) and electrostatic force microscopy (EFM) both in dark and under photo-excitation to study the impact of plasmonic generated thermal energy on the the electrostatic potential profile of its surrounding semiconductor. We investigated composite gold nanoparticle/semiconductor light-harvesting device geometry and obtained experimental evidence that such effects as plasmoelectric, thermoelectric, and hot carrier–induced are negligible compared to potentials generated in the semiconductor material itself. As the semiconductor we utilized poly (3,4-ethylenedioxythiophene) (PEDOT): poly (styrenesulfonate) (PSS) which is successfully used in devices such as photovoltaics as the buffer layer due to its high work function, conductivity and transparency[11,12].

Current profiles of the 50-nm Au colloids on indium tin oxide (ITO)/glass covered by 40 nm of PEDOT: PSS[13] measured using C-AFM are shown in Fig. 1. C-AFM has been widely used to measure the local electronic properties of various semiconductors. However, we are not aware of any reports describing the use of C-AFM to show local changes in current distribution around plasmonic metal nanoparticles upon photoexcitation. Figure 1(a) demonstrates the AFM topography image of the AuNPs sample. The residual current of C-AFM controllers is measured when no biasing voltage is applied and is offset accordingly in subsequent measurements (Fig. 1(b)). Applying a small biasing voltage (250mV) between a conductive tip and the bottom ITO electrode, the current flow through the AuNP embedded PEDOT:PSS layer is locally measured at nanoscale simultaneously with the surface topography using a platinum (Pt) coated conducting tip (Fig. 2(c)). A control film of bare PEDOT: PSS of the same thickness is used to compare the influence from plasmonic excitation. In absence of photoexcitation, films behave Ohmic and the nanoscale current map shows a relative distribution of current active regions. Within the AuNP embedded films most of the current passes through the particles and show as brighter regions. Total current measured within the image follows a direct Ohmic relationship with the applied biasing voltage. The control film does not show any morphological change or increase in total current upon photoexcitation up to the maximum available optical power. Since PEDOT:PSS has very little or no absorption at the working excitation wavelength (532nm), we do not see any significant heating or thermal damage within the scan period and power range. However, upon photoexcitation, the Ohmic relationship changes with both increasing optical power and biasing voltage. At any particular applied voltage with increasing optical excitation power the PEDOT:PSS region surrounding the Au particles becomes more conducting and at about $10^6 W/m^2$ incident power the relative current distribution changes. At this optical power, the PEDOT: PSS layer appears to carry more current than the gold particles (Fig. 2(d)). Figure 2 shows the measured current profiles under different illumination power. As the power increases, current distribution shifts towards PEDOT:PSS layer; reducing current density on top of Au nanoparticles. Total current measured at any biasing voltage increases 3–5% within the power range $10^4$ to $10^8$ W/m$^2$. Contact potential difference measured before and after pC-AFM measurements using Pt and Au coated tips did not show any variation and indicates that the change in current measured using pC-AFM is not due to degeneration or degradation of tip material.

      Within the optical, electrical, and thermal regimes, the following six effects are of importance in describing the observed change in current density: i) molecular vibration limited mobility increase within PEDOT:PSS layer due to photothermal energy transfer from plasmonic nanoparticles, ii) enhanced local mobility due to local plasmonic E-field interaction, iii) plasmoelectric effect, iv) tip - sample interaction and change in contact potential difference during photoexcitation, v) Seebeck effect on C-AFM tip due to a large bias voltage gradient present across nanoparticles, and lastly vi) increased softness within the polymer layer in the immediate vicinity of the nanoparticle due to photothermal energy transfer. While the first three effects will be extensively studied in this work, we comment briefly on other mechanisms.

      Although total current increase is not significant, the small increase could result from softening of the polymer around the plasmonic nanoparticles due to thermal heating. Another cause of large change in pC-AFM signals in the electrical measurements is the fact that large temperature gradients could lead to spurious signals through the Seebeck effect. The most straightforward way to evaluate the significance of these potential effects is to measure the steady-state change in local temperature induced by our focused laser spot. Hinz and co-workers[14] provide a thorough experimental treatment of how the nano-indentation characteristics of thin polymer films change as a function of temperature. In particular the energy dissipated in the film by a penetrating cantilever, termed the "hysteresis energy", provides a useful correlation with temperature. Warming the polymer promotes elastic (reversible) deformation over plastic (irreversible) deformation. Thus, the energy dissipated in the film from an impinging AFM tip decreases with increasing temperature. Same Pt coated tip was used for indentation and hysteresis energy measurements shown in Fig. 2. To ensure that the tip – laser spot relative positions does not change the laser spot was aligned by maximizing surface potential microscopy feedback voltage and a high gain feedback locked loop ensures minimum drift from the peizo scanner. Our measurements show a

significant increase in heat induced softness around the nanoparticles during photoexcitation and as much as 14 – 18K increase in temperature over $10^4$ to $10^7$ W/m$^2$ photoexcitation power. Whereas this proves beyond doubt that photothermal effect from plasmonic gold nanoparticles influences photoconductive currents significantly, this does not rule out contribution from E-field enhancements.

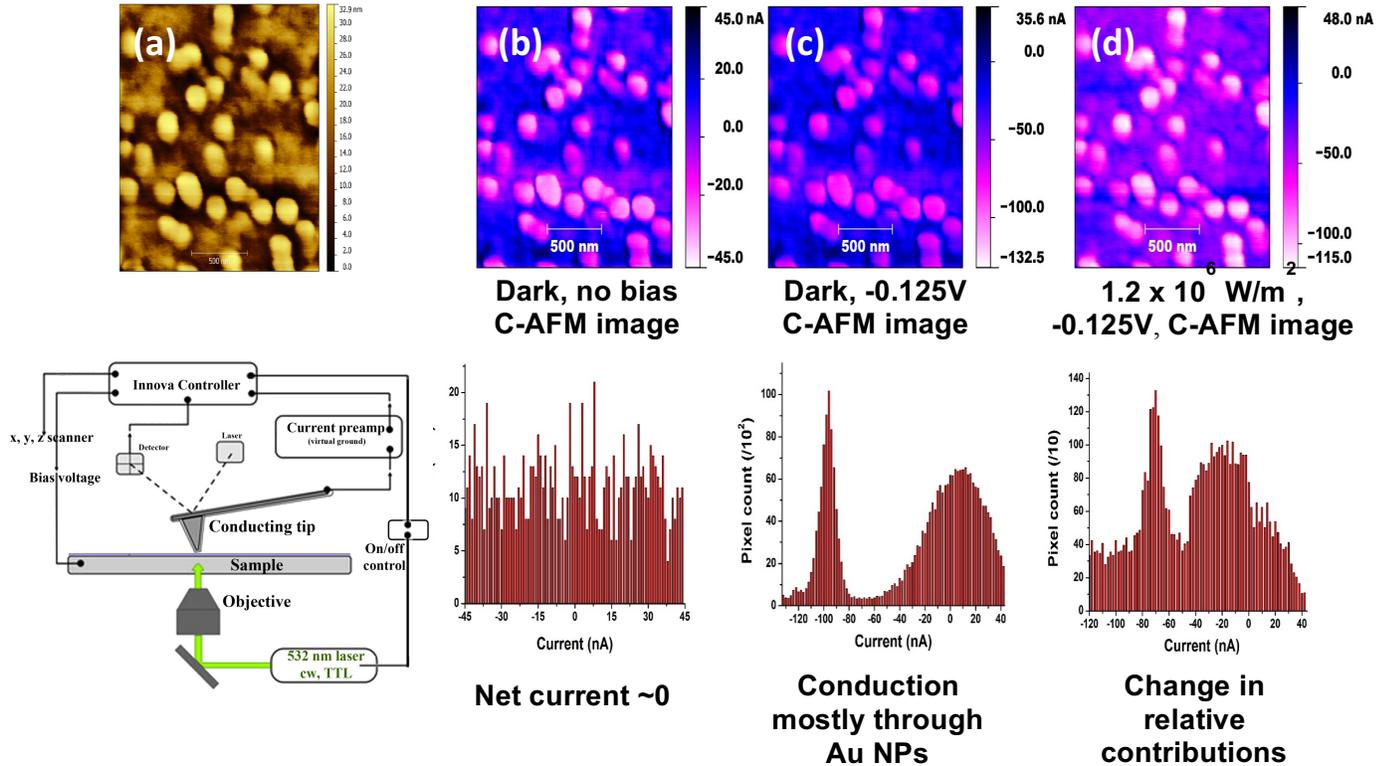

Figure 1(a) A topographic map of Au nanoparticles covalently attached to ITO surface; the particles (diameter ~50nm) are partially covered by surrounding layer of PEDOT:PSS, (b) a C-AFM image of the same region; residual image with no applied bias is from the leakage current from AFM's controller. The current measured from the image was used as offset for the ADC / DAC controllers for AFM, (c) a C-AFM image after applying bias voltage which alters the relative distribution of current on AuNPs and PEDOT:PSS, (d) photoexcited conducting AFM image showing relative current distribution over the entire sample. Current distribution histograms from the C-AFM and pC-AFM measurements are shown below their current profile.

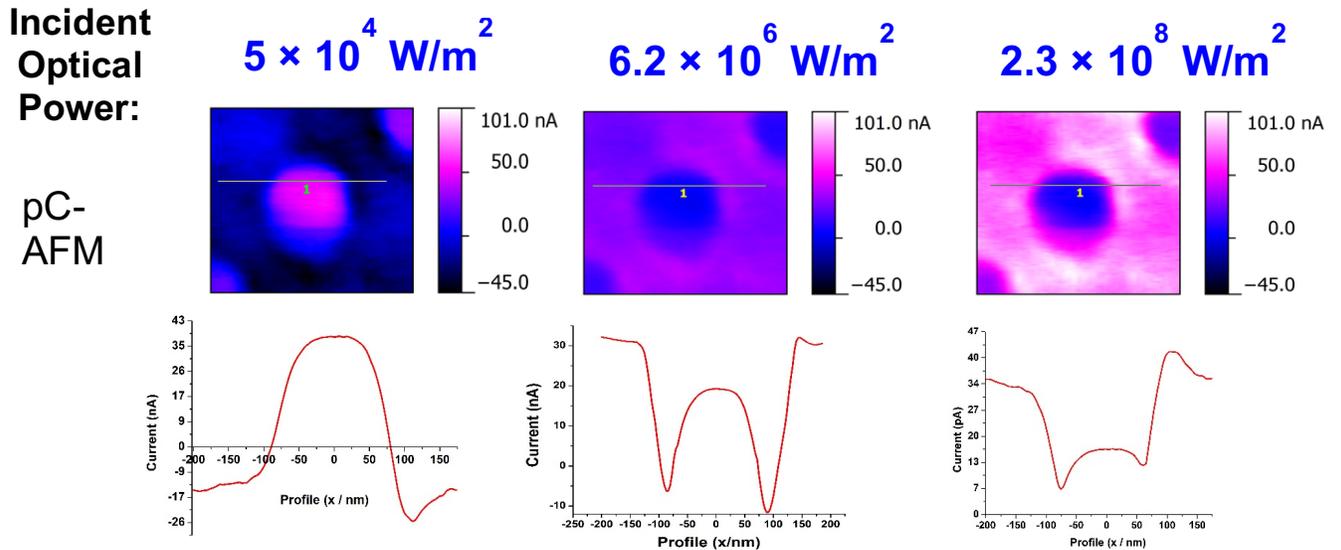

**pC-AFM current distribution under + 0.13V across the profile scan line for different incident optical power**

Figure 2. pC-AFM measured current profiles of a single AuNP surrounded by PEDOT: PSS for different excitation power.

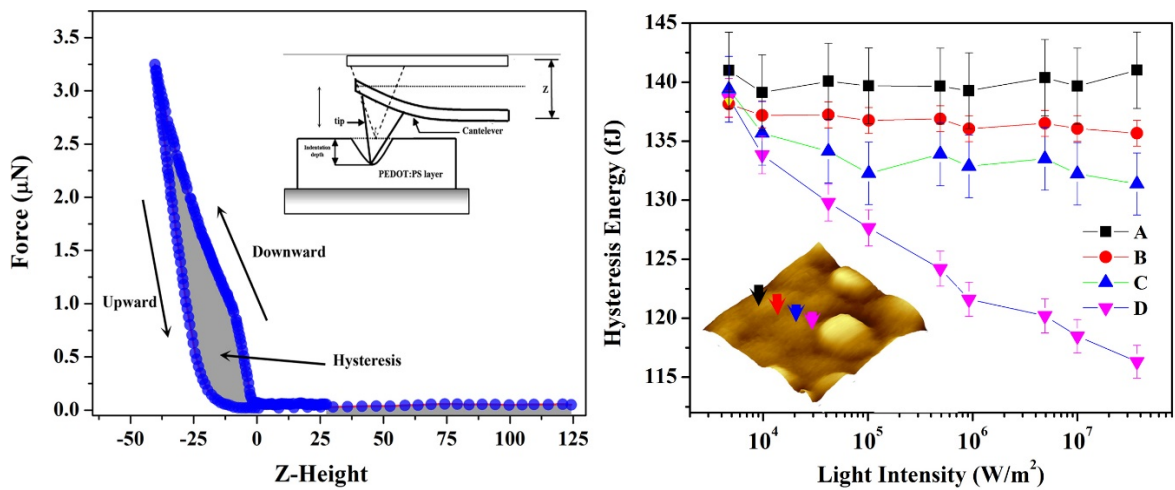

Figure 3. Left: Example of a typical force-distance curve recorded on an indentation experiment using Pt coated tip on Au NP embedded in PEDOT:PSS showing the downward (indentation) and upward (disengage) directions. The grey solid illustrates the hysteresis energy – the area between the two curves. Right: Hysteresis energy measured as a function of illumination intensity away from plasmonic particle. Time gap between each measurement is roughly 20 seconds, during which time the sample was under continuous illumination. All error bars represent the standard deviation of the mean.

To explain the experimentally observed changes in current distribution we calculated the temperature and electric field profiles of the photo-excited spherical 50-nm-diameter Au particle on an ITO/glass substrate covered by 40 nm PEDOT: PSS using COMSOL multi-physics simulations. Figure 4(a) shows a cross-section of the normalized E-field intensity distribution of AuNPs inside the buffer layer after photo-excitation at 473nm. In this case, the localized surface plasmon resonance (LSPR) of the NPs get excited and the peak field intensity at the surface of the nanoparticles is about two orders of magnitude larger than that of the incident field. Furthermore, the photoexcitation of the gold nanoparticle free electrons is followed by their cooling back to equilibrium by energy exchange with lattice phonons at the rate of ~1 ps, heating up the nanoparticle. At slower rates, the lattice cools via phonon-phonon processes (~100 ps) leading to heating of the medium surrounding the nanoparticle. Fig. 4(b) shows the modeled normalized temperature distribution of our optically excited sample. The temperature distribution can be described by the heat transfer equation

$$\rho(r)c(r)\frac{\partial T(r,t)}{\partial t} = \nabla \cdot \left(k(r)\nabla T(r,t)\right) + P(r,t) \qquad (1)$$

where t and r are time and position coordinate, T(r,t) is the local temperature, P(r,t) is the external power source density, $\rho(r)$, $c(r)$ and $k(r)$ are the mass density, specific heat, and the thermal conductivity of the material respectively. In the steady state regime, the thermal diffusion equation reduces to

$$\nabla \cdot \left(k(r)\nabla T(r,t)\right) = -P(r) \qquad (2)$$

where for a plane wave of frequency $\omega$ and an amplitude of E, P(r) would be (in $W/m^3$)

$$P(r) = \frac{\omega}{2}\varepsilon_0 Im(\varepsilon)|E|^2 \qquad (3)$$

In the COMSOL simulation, we assume that the heat power sources are mostly due to the optical absorption of AuNPs and the absorption of other layers is negligible. Thus, we locate power sources calculated from Eqn. 3 inside the AuNPs and simulate the thermal diffusion around them. The plasmonic excitation of AuNPs increases the temperature of the particle's surface up to 325K (the simulations are done at 293K). The generated heat diffuses away from the MNPs surface and increases the temperature of the surrounding medium.

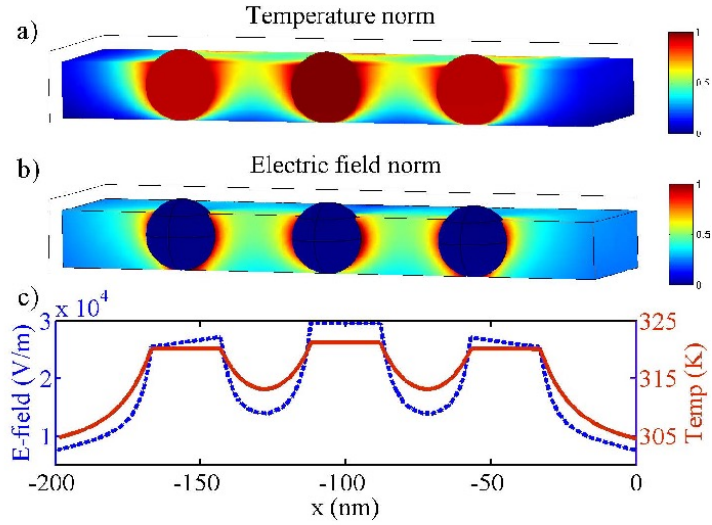

Figure 4. Modeled a) normalized temperature and b) normalized E-field intensity profiles of AuNPs in PEDOT: PSS when illuminated at λ = 473nm, c) line plots of (a) and (b), blue dotted and red solid lines show E-field and temperature distribution; respectively.

Next, we used the simulation results to calculate the conductivity of PEDOT: PSS which obeys the hopping conduction law. The conductivities in the plane ($\sigma_{||}$) and normal to the plane ($\sigma_\perp$) of PEDOT: PSS thin film are different which can be explained by its morphological structure. The electric field/temperature dependence of PEDOT: PSS conductivity at modest fields can be written as[15,16]

$$\sigma_{||}(E_{||}, T) = \sigma_{0,||}(0,T)\exp\left(0.17\frac{eE_{||}L_{||}}{k_B T}\right) \quad (4)$$

$$\sigma_\perp(E_\perp, T) = \sigma_{0,\perp}(0,T)\left[1 + \frac{1}{6}\left(\frac{eE_\perp L_\perp}{k_B T}\right)^2\right] \quad (5)$$

where $\sigma_0(0,T)$ is the DC conductivity, T is the temperature, e is the electron charge, E and L are the applied electric field and the characteristic hopping length in the field direction, respectively and $k_B$ is the Boltzman constant. The conductivity of PEDOT: PSS before and after AuNPs SPR excitation was calculated from Eqn. 4,5. The photo-excitation of AuNPs increases the conductivity of PEDOT: PSS for both in plane and normal directions ($\sigma_{||}$ and $\sigma_\perp$). The increase in $\sigma_{||}$ (~28.5%) is much larger than $\sigma_\perp$ (~2.4%) which is expected due to the morphology of spin-cast PEDOT: PSS films[12,15]. Figure 5 demonstrates the contributions of both the temperature and E-field intensity increase of photo-excited AuNPs to the conductivity increase of PEDOT: PSS. The in-plane temperature change and the E-field enhancement of the AuNPs contribute to a ~28.2% and 0.24% increase in the conductivity of PEDOT: PSS, respectively. Normal to the film these numbers are ~2.4% and ~7.8E-6%.

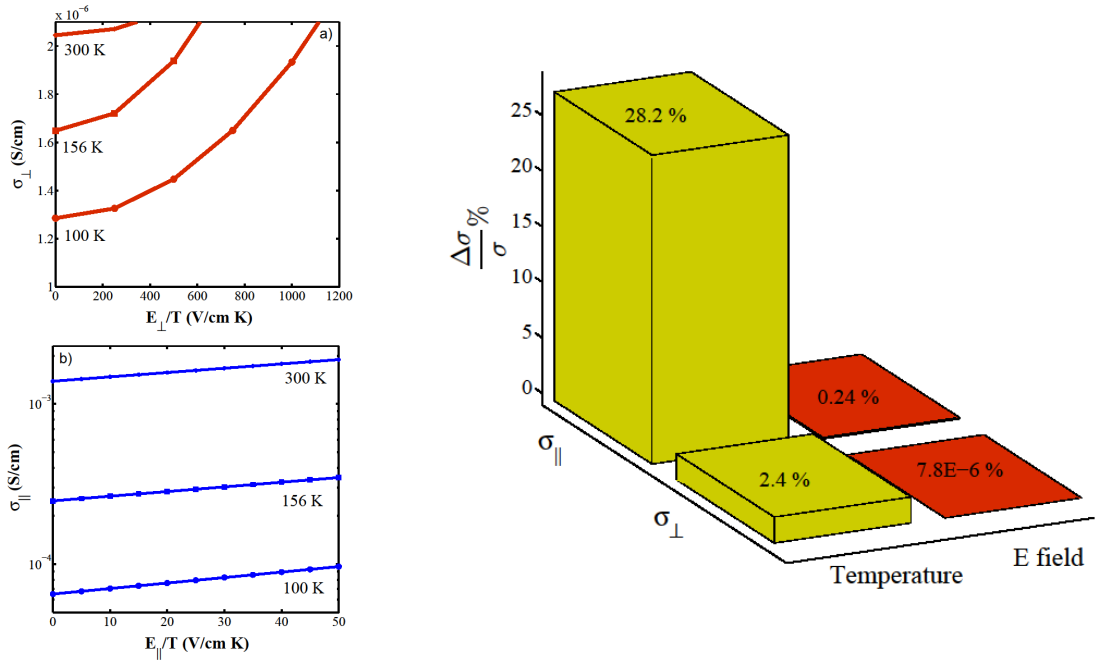

Figure 5. PEDOT: PSS conductivity (a) normal and (b) in the plane versus $E/T$ $(V/cm\ K)$ for three temperatures of 100, 156, and 300 K. (c) contributions of the temperature and E-field intensity of excited AuNPs SPR to the overall conductivity increase of PEDOT: PSS in both lateral and vertical directions.

The model on the PEDOT:PSS conductivity supports the key factors of molecular vibration limited mobility increase within PEDOT:PSS layer due to photothermal energy transfer from plasmonic nanoparticles and enhanced local mobility due to local plasmonic E-field interaction. To demonstrate the strength of this potential versus plasmoelectric mechanism, we probed the local static surface potential of the sample at room temperature by extensive EFM, photoexcited EFM experiments in dark and under illumination. The sample is first attached onto the EFM holder and the PEDOT:PSS layer on the edge of the sample is scraped. Then, we connect the ITO and the holder to provide a bias voltage of 1V. In this measurement setup, the cantilever is antimony (n) doped silicon with an average force constant of 3 N.m$^{-1}$ and resonance frequency of 80 kHz. First, the surface topography of the sample is imaged (AFM mode). Then, while a bias voltage is applied to the sample the cantilever retraces the topography with a fixed distance of 20nm above the sample (EFM mode) where its oscillation is only sensitive to the long-range electrostatic forces ($F \propto z^2$) but not the shorter-range van der Waals interactions ($F \propto z^6$). Figure 5(a) and (b) show the topography and phase of the sample measured using AFM and (c) and (d) show the dark and photo-excited EFM results of the sample respectively. The photo-excitation of the sample decreases the surface potential contrast between the AuNPs and PEDOT:PSS by ~160 mV. This change suggests that the SPR excitation of AuNPs decreases the surface charge accumulation by possibly altering the conductivity of PEDOT:PSS. A plasmoelectric effect can be estimated from results shown in Fig. 2 where the potential difference between the AuNP and PEDOT:PSS top surfaces is 2E4 $V/m \times$ 10 nm=0.2 mV which is around three orders of magnitude weaker than the observed potential.

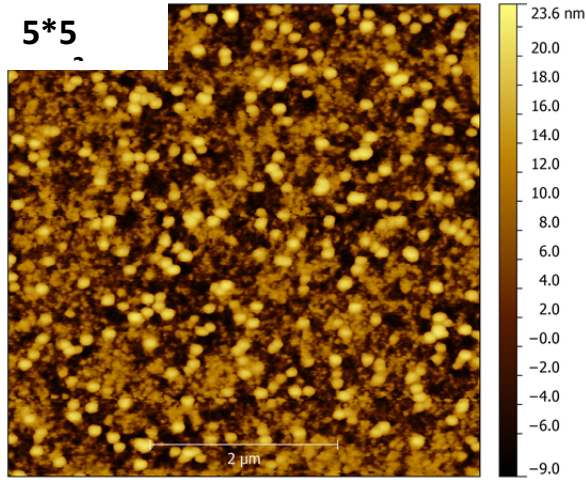
**Topo map**

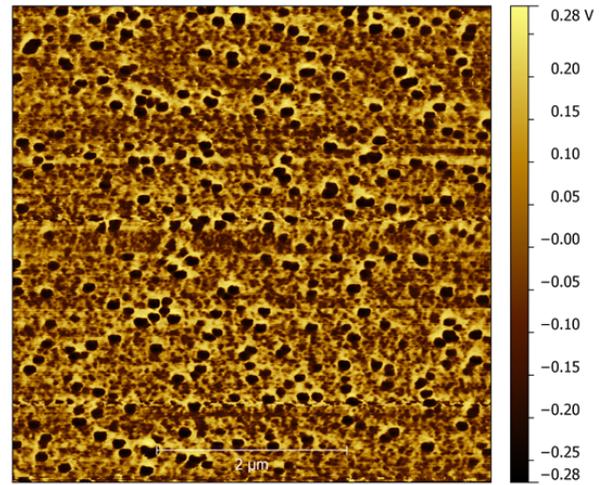
**EFM map**

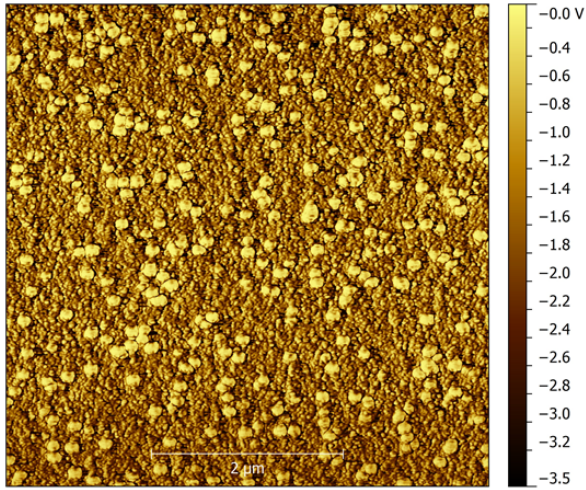
**Phase**

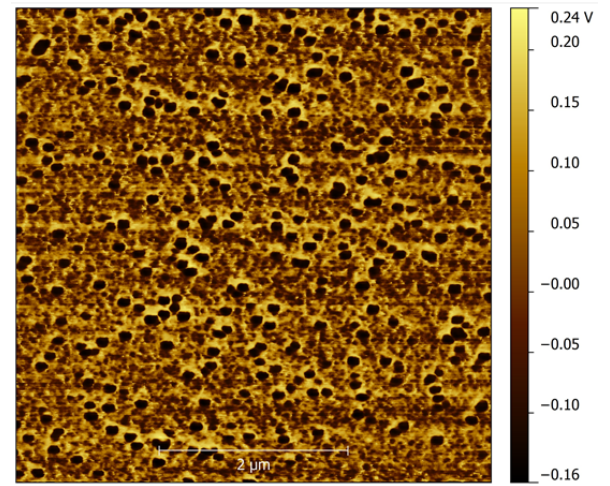
**PEEFM map**

Figure 6. (A) AFM surface topography, (B) AFM phase image, (C) EFM electrostatic potential map with a bias voltage of 1V applied between the tip and the ITO substrate and the dark lift of 20nm, (D) PEEFM electrostatic potential map with $10^5 W/m^2$ laser intensity.

1. Atwater, H. A. & Polman, A. Plasmonics for improved photovoltaic devices. *Nat. Mater.* **9,** 865 (2010).

2. Shahin, S., Gangopadhyay, P. & Norwood, R. A. Ultrathin organic bulk heterojunction solar cells: Plasmon enhanced performance using Au nanoparticles. *Appl. Phys. Lett.* **101,** (2012).

3. Qiao, L. *et al.* Localized surface plasmon resonance enhanced organic solar cell with gold nanospheres. *Appl. Energy* **88,** 848–852 (2011).

4. Chen, F. C. *et al.* No Title. *Appl. Phys. Lett.* **95,** 13305 (2009).

5. Wang, D. H. *et al.* Enhanced light harvesting in bulk heterojunction photovoltaic devices with shape-controlled Ag nanomaterials: Ag nanoparticles versus Ag nanoplates. *RSC Advances* **2,** 7268 (2012).

6. Brongersma, M. L., Halas, N. J. & Nordlander, P. Plasmon-induced hot carrier science and technology. *Nat. Nanotechnol.* **10,** 25–34 (2015).

7. Clavero, C. Plasmon-induced hot-electron generation at nanoparticle/metal-oxide interfaces for photovoltaic and photocatalytic devices. *Nat. Photonics* **8,** 95–103 (2014).

8. Govorov, A. O. & Richardson, H. H. Generating heat with metal nanoparticles. *Nano Today* **2,** 30–38 (2007).

9. Baffou, G., Quidant, R. & de Abajo, F. J. Nanoscale Control of Optical Heating in Complex Plasmonic Systems. *ACS Nano* **4,** 709–716 (2010).

10. Sheldon, M. T., van de Groep, J., Brown, A. M., Polman, A. & Atwater, H. A. Plasmoelectric potentials in metal nanostructures. *Science (80-. )*. **346,** 828–831 (2014).

11. Nardes, A. M. *et al.* Microscopic understanding of the anisotropic conductivity of PEDOT:PSS thin films. *Adv. Mater.* **19,** 1196–1200 (2007).

12. Nardes, A. M. *et al.* Microscopic Understanding of the Anisotropic Conductivity of PEDOT:PSS Thin Films. *Adv. Mater.* **19,** 1196–1200 (2007).

13. Shahin, S., Gangopadhyay, P. & Norwood, R. A. Ultrathin organic bulk heterojunction solar cells: Plasmon enhanced performance using Au nanoparticles. *Appl. Phys. Lett.* **101,** 053109 (2012).

14. Hinz, M. *et al.* Temperature dependent nano indentation of thin polymer films with the scanning force microscope. *Eur. Polym. J.* **40,** 957–964 (2004).

15. Ionescu-Zanetti, C., Mechler, A., Carter, S. A. & Lal, R. Semiconductive Polymer Blends: Correlating Structure with Transport Properties at the Nanoscale. *Adv. Mater.* **16,** 385–389 (2004).

16. Nardes, A. M., Kemerink, M. & Janssen, R. A. J. Anisotropic hopping conduction in spin-coated PEDOT:PSS thin films. *Phys. Rev. B* **76,** 85208 (2007).

# Plasmonic Induced Potential in Metal-Semiconductor Composits

*Shiva Shahin\*, Palash Gangopadhyay, and Robert A. Norwood*

*College of Optical Sciences, University of Arizona, Tucson, AZ, USA*

[*shshahin@ucsd.edu](mailto:shshahin@ucsd.edu)

# Supplementary Materials:

**EXPERIMENTAL**

A.   Atomic Force Microscopy:  Atomic (AFM), conductive (C-AFM) and electric force (EFM) microscopy measurements were carried out on a Veeco Innova Atomic Force Microscope. Samples were mounted on a non-polarizing beam splitter fixed on top of the xyz piezo scanner.  A small piece of wire were soldered on top of ITO surface to apply voltage.  Conducting tips platinum coated Olympus OSCM-PT-R3 (spring constants ~2 $Nm^{-1}$, quality factor ~ and resonant frequency of ~70 KHz) and Gold coated µMash CR-AU 18 series (spring constants ~2.8 $Nm^{-1}$, quality factor ~ and resonant frequency of ~75 KHz) were used for electrostatic force microscopy (EFM) whereas platinum-iridium coated Bruker SCM-PIC (spring constants ~0.2 $Nm^{-1}$) was used for contact mode imaging and pc- and C-AFM.  Innova control was programmed to apply voltages to the sample via substrate and to collect all images using two pass method.  For all photoexcited measurements the laser was synchronized to be put-on at the start of the backward pass whereas the forward pass was used to capture topographic image and C-AFM images.  A delay of 300 mS was set at the start and end of backward pass to stabilize laser and any effect of photoexcitation. A laser diode working at 532nm (Lasermate GMSLF-532-20E) was collimated after a spatial filter with 500µm aperture and focused at the top of ITO using a 20x objective.  The area illuminated was ~250 µm in diameter.  Continuously variable neutral density (ND) filter was inserted to keep charging rate within 1000Hz.  AFM lateral dimensions and depth were calibrated using calibration references supplied by Bruker, APCS-0001 (100nm depth and 1µm pitch) and NIST traceable STS3-1000P (100nm step and 3/10/20µm pitch) whereas 1kΩ and 1M kΩ head mounted oxide resistors were used to calibrate digital-to-analogue and analogue-to-digital signal converters and to check linearity of preamp of C-AFM module.

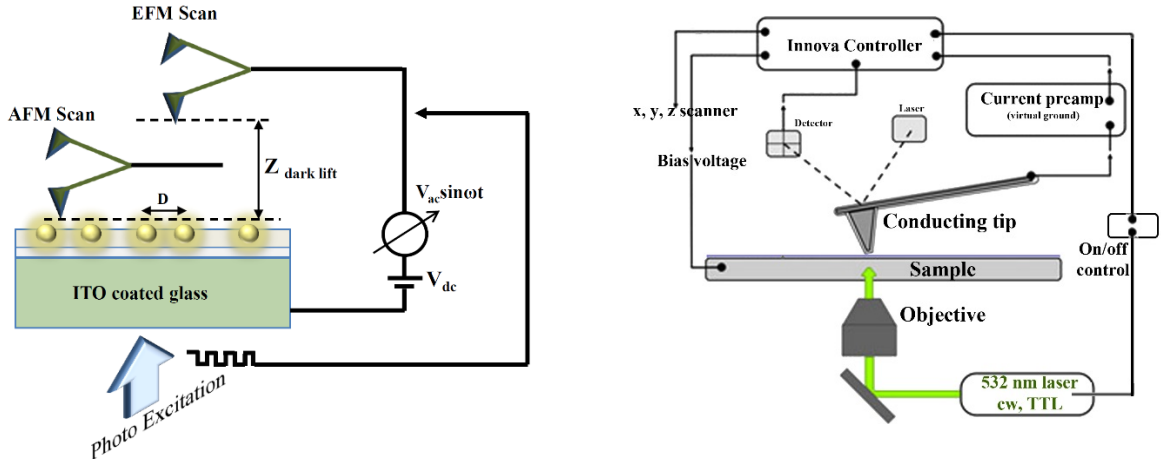

Figure S1. Left: Schematic diagram of the photoexcited electrostatic force microscopy (PEEFM) setup; Right: Schematic diagram of photoconductive AFM (pc-AFM) setup used in this study.

B.  Sample Preparation: The plasmonic structure used in this work is fabricated according to our previous report[21] where clean ITO coated substrates are activated in a dilute HCl solution (5% v/v in water) for 5 seconds, followed by copious rinsing with DI water and drying in a nitrogen stream. Afterward, the activated ITO samples are immersed in dilute $N^1$-(3 trimethoxysilylpropyl)diethylenetriamine (DETA) (3% v/v in diluted in ethanol) and held at 85°C for an hour. Finally, the substrates are dipped into the gold nanosphere solution for 48 hours. To modify the silanized ITO, a thin layer (40nm) of PEDOT:PSS was spun coat on samples in two steps of 2700 and 4000 rpm each for 1 minute. Then the samples are annealed at 140°C for 10 minutes.

## Hopping conduction law for PEDOT: PSS

PEDOT: PSS is a mixture of PEDOT, a semiconducting polymer that becomes conductive when poled, and PSS, a weak ionic conductor[23]. The conductivity of PEDOT: PSS obeys the hopping conduction law. A study on the morphology of thin films of PEDOT: PSS shows that the conductive grains of PEDOT or a mixture of PEDOT and PSS are separated by insulating shells of PSS[24,25]. The conductivity in the plane of a thin film of PEDOT: PSS is about three orders of magnitude larger than the conductivity normal to the plane which can be explained by its morphological structure. The temperature dependencies of the conductivity ($\sigma$) in the Ohmic or low electric field regime for spin-coated PEDOT: PSS thin films is shown in (3)

$$\sigma(T) = \sigma_0 \exp\left[-\left(\frac{T_0}{T}\right)^\alpha\right] \qquad (1)$$

where $T_0$ is a material-dependent parameter and the exponent $\sigma = 1/(1 + D)$ is taken as variable range hopping (VRH) parameter in D dimensions. In this work we use $\alpha = 0.25$, $T_0 = 3.2\times10^6 K$, $\sigma_0 = 24.7\ S/cm$ for the conductivity in the plane and $\alpha = 0.81$, $T0 = 70K$, $\sigma_0 = 2.6\times10^{-6}\ S/cm$ normal to the plane[26]. The electric field dependencies of PEDOT: PSS conductivity at modest fields $eE_{||}L_{||} / k_BT < 1$ can be written as

$$\sigma_{||}(E_{||}, T) = \sigma_{0,||}(0, T)\exp\left(0.17 \frac{eE_{||} L_{||}}{k_B T}\right) \quad (2)$$

$$\sigma_{\perp}(E_{\perp}, T) = \sigma_{0,\perp}(0, T)\left[1 + \frac{1}{6}\left(\frac{eE_{\perp} L_{\perp}}{k_B T}\right)^2\right] \quad (3)$$

where $\sigma_0(0, T)$ is the DC conductivity, e is the electron charge, E and L are the applied electric field and the characteristic hopping length in the field direction, respectively and $k_B$ is the Boltzman constant.